# DOMINANT BLOCK GUIDED OPTIMAL CACHE SIZE ESTIMATION TO MAXIMIZE IPC OF EMBEDDED SOFTWARE

Rajendra Patel[1] and Arvind Rajawat[2]

[1,2]Department of Electronics and Communication Engineering,
Maulana Azad National Institute of Technology (MANIT), Bhopal, India

## ABSTRACT

*Embedded system software is highly constrained from performance, memory footprint, energy consumption and implementing cost view point. It is always desirable to obtain better Instructions per Cycle (IPC). Instruction cache has major contribution in improving IPC. Cache memories are realized on the same chip where the processor is running. This considerably increases the system cost as well. Hence, it is required to maintain a trade-off between cache sizes and performance improvement offered. Determining the number of cache lines and size of cache line are important parameters for cache designing. The design space for cache is quite large. It is time taking to execute the given application with different cache sizes on an instruction set simulator (ISS) to figure out the optimal cache size. In this paper, a technique is proposed to identify a number of cache lines and cache line size for the L1 instruction cache that will offer best or nearly best IPC. Cache size is derived, at a higher abstraction level, from basic block analysis in the Low Level Virtual Machine (LLVM) environment. The cache size estimated from the LLVM environment is cross validated by simulating the set of benchmark applications with different cache sizes in SimpleScalar's out-of-order simulator. The proposed method seems to be superior in terms of estimation accuracy and/or estimation time as compared to the existing methods for estimation of optimal cache size parameters (cache line size, number of cache lines).*

## 1. INTRODUCTION

Fast design space exploration is essential in order to achieve shorter time-to-market with reasonably low cost of design. Memory utilization, Energy Consumption, Performance on Hardware and Software, Communication Overhead are a few design constraints that need to be profiled and extracted from the design description. A survey on profilers for embedded systems is presented in [3]. Profilers are required to extract such parameters from high-level software descriptions. Such profiled numerical features greatly help to narrow down the design space for detailed exploration at lower levels.

Efficient embedded systems need to be highly configurable as per application requirement, starting from functional units of the data-path, register files, control-path, cache memory, etc. The cache memory is an important part of any medium to higher level embedded system. It is always a design challenge to choose the optimal cache size, which can offer best IPC. This helps to realize a design with better performance versus cost. Cache is fast in response but at the same time it is one of the main parts of the embedded system which consumes lots of energy as discussed by Alexandra [1] and Andhi et al. [5]. For example, the instruction cache of ARM 920TTM consumes 25% of the total power as reported in [1]. For battery operated embedded systems, energy consumption is one of the tightly constrained design parameters. For such applications optimal cache size selection leads to lower energy consumption without compromising with the performance.

The cache designing involves determining cache line size, number of cache lines required, associativity, replacement and write-back policy. In this work two important parameters, cache line size and numbers of cache lines, are estimated for a given application in order to achieve maximized IPC. It is noted that in general the performance of a system improves with an increase in the cache size. However, such performance improvement seizes after certain cache size as shown by Abhijit et al. [4]. For a fixed size (in number of bytes) of cache, different combinations of the cache line size and total number of cache lines are possible. It is required to simulate all such combinations and identify an optimal cache line size and the number of cache-lines combination. For a given size of cache, with different combination of the cache line size and number of cache lines, variation in performance is observed as reported in Table 4, Section 4. Hence, it is required to tune the cache size with the correct combination of line size and number of lines to obtain improved IPC. The design space for selecting the right combination of line size and number of lines is large. In this paper, a method based on basic block analysis of the intermediate level code is proposed for rapid estimation of the cache line size and number of cache lines for a given application and processor.

The rest of the paper is organized as follows. Section 2 deals with the related work on the topic discussed in this paper. Section 3 discusses the proposed cache estimation technique. The results of the work done are presented in Section 4. Discussion on observations made in this work is given in Section 5. Finally Section 6 presents the concluding remarks on the work done.

## 2. RELATED WORK

Several works in the last few years have addressed the significance of cache design. Some researchers have focused on the optimal cache design to achieve less energy consumption in the cache circuits. It is also explored to minimize area usage and obtain the optimal performance for embedded processors.

Alexandra [1] has proposed a unique cache hierarchy iLP-NUCA (Instruction Low Power – Non Uniform Cache Architecture) for high performance low power embedded system. It replaces the conventional L2 cache and improves the energy-delay of the system. This cache architecture has shown significant improvement in the multi-threaded multiprocessor system.

The effect of cache and register file size is very crucial in achieving better performance of embedded processors as discussed by Mehdi et al. in [2]. Since both of these resources are on-chip, they lead to increased area and energy consumption. The improvement in performance is observed by doubling the cache and the register file size. But such improvement is not significant after certain levels. It implies that it is necessary to select an optimum size of cache and the register file to avoid selecting the larger size of these resources.

Andhi et al. [5] have suggested a method to identify the correct combination of the cache line size, number of cache lines and associativity for both, L1 data cache and instruction cache. A method is proposed which reads the large execution trace to generate cache miss rate statistics for all possible cache configurations. The one which gives the best miss rate can be selected.

However, the drawback of the approach is that it is compulsory to execute the application on a target simulator with any one cache configuration to generate the execution trace.

A unique algorithm named Software Trace Cache (STC) is proposed by Alex et al. in [8]. The algorithm is targeted to layout the instruction basic blocks, in such a way that it minimizes the instruction cache miss rate. The STC algorithm organizes basic blocks into chains trying to make sequentially executed basic blocks reside in contiguous memory locations, and then maps such chains in memory to minimize conflict misses in the frequently executed sections of the program. The proposed approach increases both spatial and temporal locality to minimize the miss rate. The drawback of the approach is that considerable changes are required at the compiler level. This increases the design space exploration time.

Bartolini et al. [6] have proposed an instruction cache optimizing technique to achieve better performance with minimum cache size. The approach exploits Cache Aware Code Allocation Technique (CAT) which allocates the memory to different segments of the code to improve localities. This approach is also very much similar to the method discussed by Alex et al. in [8]. Abhijit et al. in [4] have presented an efficient technique to estimate the processor performance by examining the intermediate level code and approximating the machine level code from it. The methodology is validated on ARM and PISA architectures. The LLVM development environment [13] is used for development of this technique. It is assumed in this exploration that a large cache is available for instruction as well as data. IPC was estimated based on the basic block analysis done at intermediate level in LLVM.

Arijit et al. in [9] have proposed an analytical approach to come up with a cache configuration for a given application. An algorithm that tunes the cache parameters, cache lines and associativity, has been proposed. A fast convergence to a desired solution has been claimed compared to traditional design-simulate-analyze approach.

Marc et al. in [7] have discussed cache designing methodology based on locality analysis performed on the execution trace of the application. This approach is more suitable to system whose software is not changing throughout its lifetime, like embedded software.

Yau-Tsun et al. [11] have proposed a method to estimate the worst case execution time (WCET) for given application implemented on a heterogeneous system using cache based memory sub-system. The WCET is highly dependent on the nature of input data and the initial state of the system. The actual WCET can be obtained only after a large number of simulations which is impractical. Hence, a static analysis of the program is performed to arrive at tightly bounded WCET which must lie within accepted limits of WCET. Integer linear programming is used to propose this technique.

## 3. CACHE SIZE ESTIMATION TECHNIQUE

Major steps in the estimation of the instruction cache are shown in Figure 1. The proposed technique exploits features of two open source tools, LLVM 3.1 [12] and SimpleScalar 2.0 [13]. LLVM is a compiler framework designed to support program analysis and transformation of programs. The application code is analyzed at an intermediate level using LLVM profiler. To carry out basic block analysis, the LLVM byte code is instrumented for edge profiling. The basic blocks and their corresponding execution frequencies are obtained by executing the instrumented code on the host.

There are different simulators available in the SimpleScalar toolset, which offers a wide range of execution speed versus execution statistics. The sim-fast is the fastest functional simulator but does not provide any statistics about the application. Whereas, sim-outorder is the slowest but provides lots of details like cache performance, execution clocks, etc. In this work, sim-outorder for ARM and PISA target processors is used to simulate IPC for different applications from MiBench [10]. To use simple scalar simulators, it is required to configure the configuration file as per the architectural features expected. To simulate applications in the SimpleScalar single issue pipeline with single ALU and Multiplier architecture is used. The output of the simulator is the IPC observed during execution of the given application. This information is used to cross validate the cache line size and number of cache lines estimated from basic block analysis for optimal IPC.

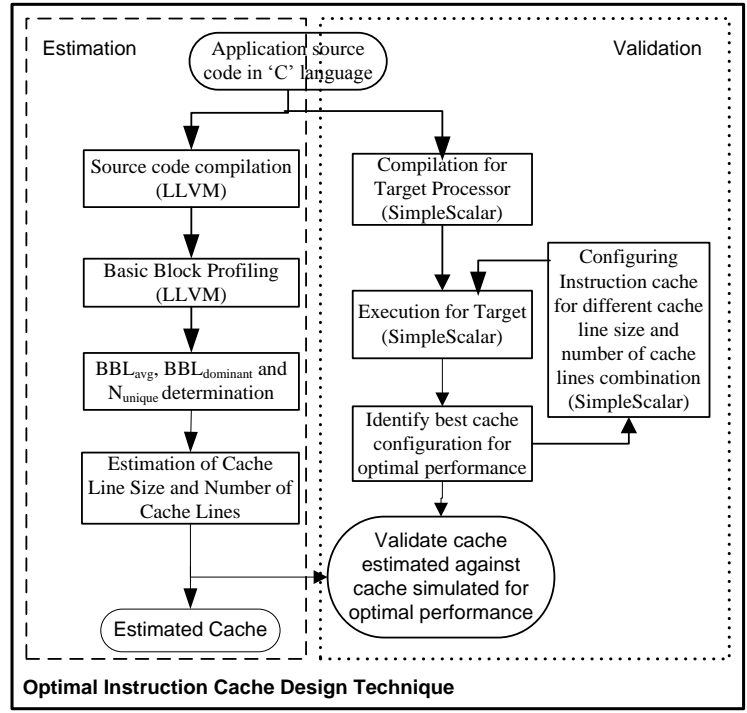

Figure 1 Instruction Cache Size Estimation Technique

The area covered with dashed line in Figure 1 represents the flow for cache size estimation and the area in dotted line represents the steps to validate the estimated cache.

### 3.1. Cache Size Determination

The cache size is determined based on cache block size (cache line size) and the number of cache lines required. It is required to note that in any system, the cache block size and required cache lines are chosen in power of 2.

Table 1 Basic Block Analysis

| Application | $N_{uniqueblocks}$ | $BL_{avg}$ | $BL_{domi}$ | $BL_{largest}$ |
|---|---|---|---|---|
| Qsort | 102 | 8 | 15 | 19 |
| Dijkstra | 49 | 5 | 9 | 26 |
| Sha | 46 | 12 | 23 | 40 |
| Rawcaudio | 32 | 7 | 17 | 40 |
| crc32 | 11 | 13 | 20 | 20 |
| fft | 65+ | 13 | 119 | 119 |

*BL stands for Block Length

## 3.2. Number of Cache Lines Determination

The basic block analysis is performed on six different benchmark applications as shown in Table 1. The number of unique basic blocks determines the required number of cache lines. Here an attempt is made to reserve one cache line for each basic block. This is the ideal assumption in which block address space is not overlapping to neighbouring cache lines. The cache lines are rounded off to upper bound as depicted in Table 2. To ensure better performance, upper bound should be selected and for reduced cache size with degradation in performance, lower bound can also be preferred if the number of lines is closer to it.

Table 2 Number of Cache Line Selection Based on Unique Blocks

| Application | qsort | dijkstra | sha | rawcaudio | crc32 | fft |
|---|---|---|---|---|---|---|
| Number of Unique Basic Blocks | 102 | 49 | 46 | 32 | 11 | 65+ Library Functions |
| Number of Cache Lines Selected | 128 | 64 | 64 | 32 | 16 | 128 |

## 3.3. Cache Block Size Determination

The size of the cache block is determined with reference to the size of the dominant block length, average block length or largest block length. Let say there are n unique basic blocks (1,2,3,…, i, …, n) in a given application. The frequency of occurrence of these blocks is f1, f2, f3, …,fi,…, fn. The length of each block is bbl1, bbl2, bbl3, …, bbli, …, bbln. Then the dominant block is the one for which product of occurrence of the block during execution and the basic block length is maximized. The block length of dominant block is mathematically defined in equation (1). The average block length is obtained by taking the weighted average of all the basic blocks in the program as presented in equation (2). The largest block length is defined as the block with maximum number of LLVM instructions.

$$bbl_{dom} = MAX(f_i * bbl_i) \quad \ldots\ldots\ldots\ldots\ldots\ldots\ldots\ldots\ldots\ldots\ldots\ldots\ldots\ldots\ldots\ldots\ldots\ldots\ldots\ldots(1)$$

$$bbl_{avg} = \frac{1}{n} * \sum_{i=1}^{n} f_i * bbl_i \quad \ldots\ldots\ldots\ldots\ldots\ldots\ldots\ldots\ldots\ldots\ldots\ldots\ldots\ldots\ldots\ldots(2)$$

Largest block length improves the possibility to hold the entire block (including the largest block) in the corresponding cache line after the first reference in the memory at the start of the execution of the block. Cache block size according to dominant block length guarantees that at least all the instructions of the dominant block and the blocks which are shorter than the dominant block are available in the cache if the miss occur at the first instruction of the block. However this block size is not sufficient to accommodate the blocks which are larger in size compared to dominant block and therefore will cause some cache misses that in turn reduces the IPC. The third approach for cache block size determination is based on average block length. The blocks which are not very frequent, in execution, also contribute to the average block length calculation. This implies the reduced size of cache block compared to dominant or largest block length. Hence, all the blocks having a block length greater than average will be distributed over multiple cache lines. This will lead to increased cache miss rate in the majority of the applications and degraded IPC.

The LLVM architecture is a RISC machine, like PISA and ARM. Hence the majority of the computation oriented LLVM instructions are mapped on single instruction of PISA or ARM [4]. Based on this ground it is assumed that the number of instructions to be executed on the PISA or ARM machine are identical to that of LLVM instructions. Each PISA instruction is 8 bytes long. Hence the average block length, dominant block length and largest block length are multiplied by International Journal of Embedded Systems and Applications (IJESA) Vol.3, No.3, September 20138 for PISA (by 4 for ARM) and rounded off to the next higher or lower value of power of 2 to decide the cache block size as shown in Table 3.

Table 3 Cache Block Size According to Basic Block Length

| Application | | ARM | | | PISA | | |
|---|---|---|---|---|---|---|---|
| | | Average | Dominant | Largest | Average | Dominant | Largest |
| qsort | $B_{length}$ | 8 | 15 | 19 | 8 | 15 | 19 |
| | CBS | 32 | 60 ≈ 64 | 76 ≈ 64 | 64 | 120 ≈ 128 | 152 ≈ 128 |
| dijkstra | $B_{length}$ | 5 | 9 | 26 | 5 | 9 | 26 |
| | CBS | 20 ≈ 16 | 36 ≈ 32 | 104 ≈ 128 | 40 ≈ 32 | 72 ≈ 64 | 208 ≈ 256 |
| sha | $B_{length}$ | 12 | 23 | 40 | 12 | 23 | 40 |
| | CBS | 48≈32 | 92≈64 | 160≈128 | 96 ≈ 64 | 184 ≈ 128 | 320 ≈ 256 |
| rawcaudio | $B_{length}$ | 7 | 17 | 40 | 7 | 17 | 40 |
| | CBS | 28≈32 | 68≈64 | 160≈128 | 56 ≈ 64 | 136 ≈ 128 | 320 ≈ 256 |
| crc32 | $B_{length}$ | 13 | 20 | 20 | 13 | 20 | 20 |
| | CBS | 52≈64 | 80≈64 | 80≈64 | 104 ≈ 128 | 160 ≈ 128 | 160 ≈ 128 |
| fft | $B_{length}$ | 13 | 119 | 119 | 13 | 119 | 119 |
| | CBS | 52≈64 | 476≈512 | 476≈512 | 104 ≈ 128 | 952 ≈ 1024 | 952 ≈ 1024 |

*CBS = Cache Block Size

In proposed technique, the cache size is determined based on basic block analysis of the application. Basic block information is obtained using LLVM profiler. The limitation of the profiler is that it can profile only that segment of the program whose source code is available. It means LLVM cannot provide block information for library functions. From this viewpoint, the proposed technique is suitable for embedded software in which entire code is user defined and is available for profiling.

## 4. RESULT

As shown in Table 4, for four different applications, the variation in IPC up to 24.64% is observed for same cache size with different arrangement of cache line size and number of cache lines. This observation acts as motivation to present a method that suggests the line size and number of cache lines combination for the cache required for given application and processor architecture.

Table 4 IPC for Different Combination of Cache Line Size and Number of Cache Lines

| Application | Cache Size (KB) | Cache line size and number of cache line combination | | | | | | % variation of IPC in Case 2 w.r.t. Case 1 |
|---|---|---|---|---|---|---|---|---|
| | | Case 1 | | | Case 2 | | | |
| | | Line size | No. of lines | IPC | Line size | No. of lines | IPC | |
| qsort | 1 | 64 | 16 | 0.2365 | 16 | 64 | 0.2266 | 4.18 |
| sha | 4 | 128 | 32 | 0.5816 | 32 | 128 | 0.6139 | 5.55 |
| dijkstra | 8 | 256 | 32 | 0.351 | 64 | 128 | 0.4375 | 24.64 |
| rawcaudio | 1 | 128 | 8 | 0.4949 | 16 | 64 | 0.4956 | 0.14 |

The utilization of the proposed technique is validated with the help of 6 different applications of the MiBench for two different processor architectures, ARM (32 bit ISA) and PISA (64 bit ISA). Basic block information is extracted as shown in Table 1 for each application. Largest Block Length (BLlargest), Average Block Length (BLavg) and Dominant Block Length (BLdomi) are calculated for all applications. According to these lengths and unique number of blocks, cache

40

block size and number of cache lines are decided as given in Table 2 and 3. All the benchmark applications are analyzed without any modification in the code except QSORT. In QSORT, library function qsort( ) is being called in the original benchmark code. The LLVM is not able to profile qsort( ) as mentioned earlier. It was leading to incorrect cache size estimation. The problem is resolved by in-lining the library function code into the application source code. This makes possible to analyze qsort( ) along with the source code. Improved cache size estimation for QSORT application is achieved by this approach. The IPC graphs are prepared by executing the same benchmarks in the SimpleScalar simulator with different combinations of cache line size and number of lines. The IPC for the cache size estimated by the proposed technique is indicated by the arrow text markers. The IPC curves of benchmarks for ARM and PISA architectures are shown in Figure 2 and 3, respectively.

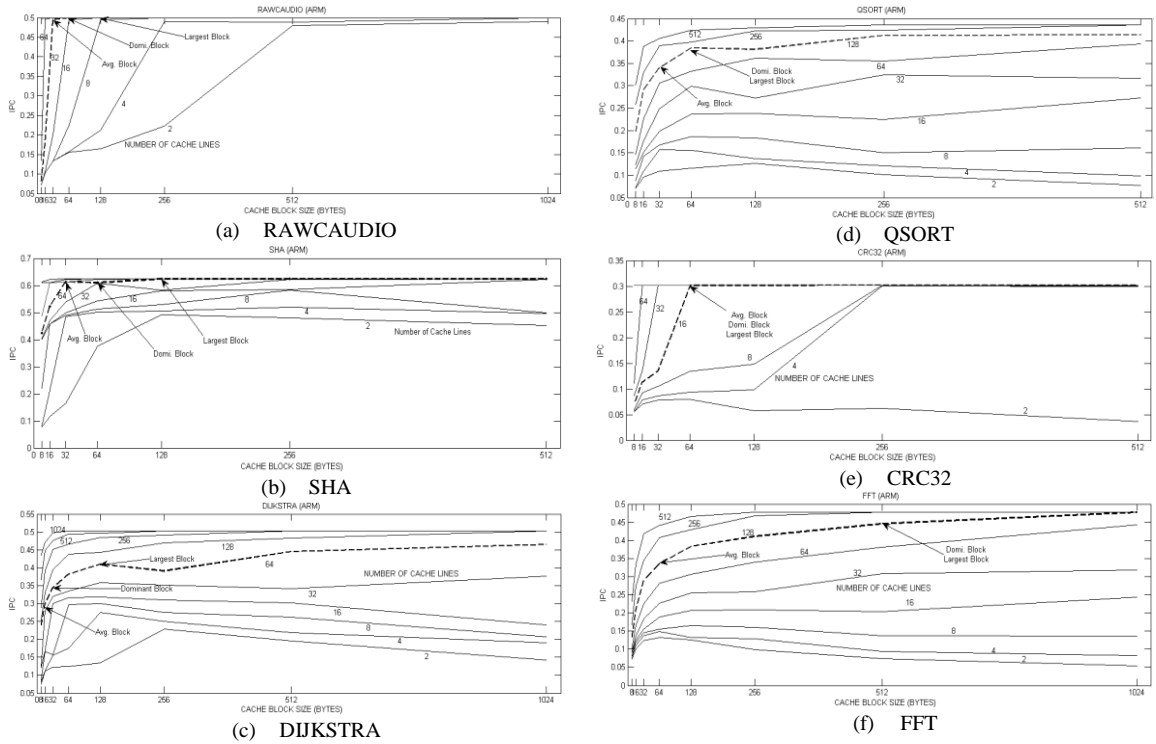

(a) RAWCAUDIO
(b) SHA
(c) DIJKSTRA
(d) QSORT
(e) CRC32
(f) FFT

Figure 2 IPC for ARM Architecture

The accuracy in estimating the highest possible IPC by choosing optimal cache size is shown in Figure 4 for ARM and PISA processors. The results represent the effectiveness of the proposed technique. In case of largest block length, average accuracy over all the benchmarks and processor architectures in IPC estimation is 94.50%. Whereas, the average accuracy for average and dominant block lengths is 81.08% and 92.39%, respectively.

The cache size estimated for various block length criteria is shown in Table 5. The average cache size estimated over all benchmarks and processor architectures for average block length, dominant block length and largest block length is 4.16, 20.16 and 23.16, respectively. The rise in cache size from average to dominant block length offers significant improvement (11.31%) in IPC. However, the minor improvement (2.11%) is noticed for largest block length over dominant block. This leads to choose dominant block length as the ideal criteria for cache size estimation in this technique.

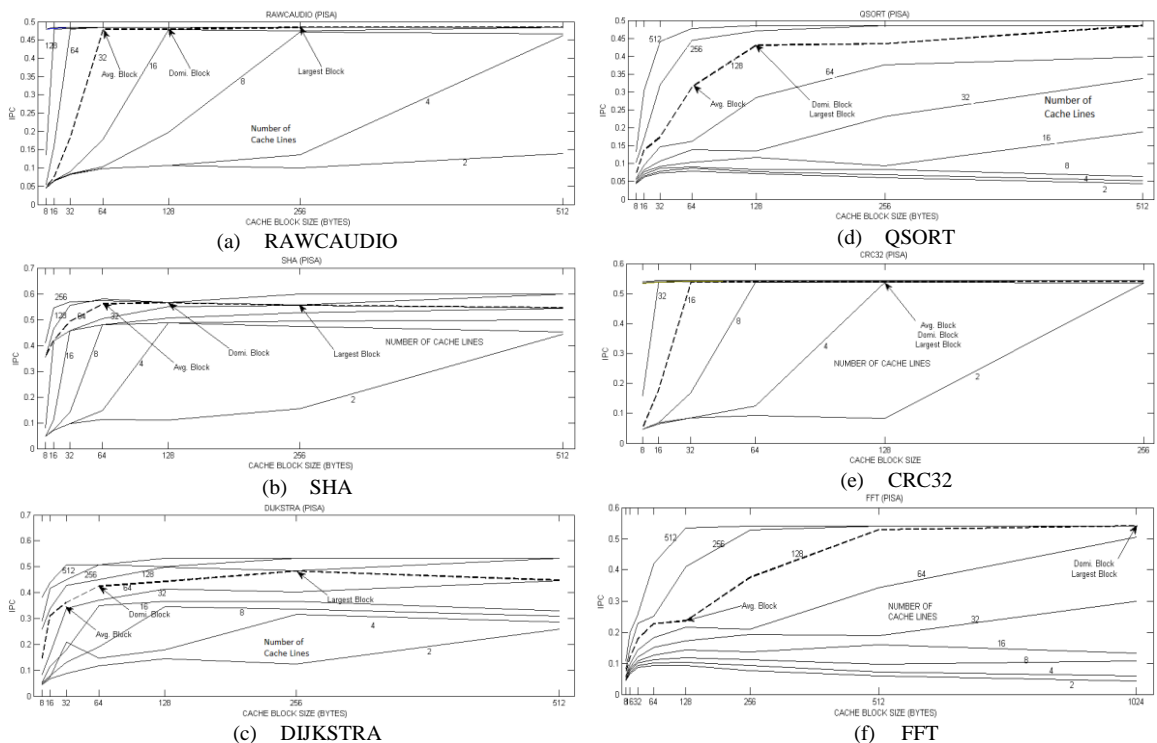

(a) RAWCAUDIO  (d) QSORT
(b) SHA  (e) CRC32
(c) DIJKSTRA  (f) FFT

Figure 3 IPC for PISA Architecture

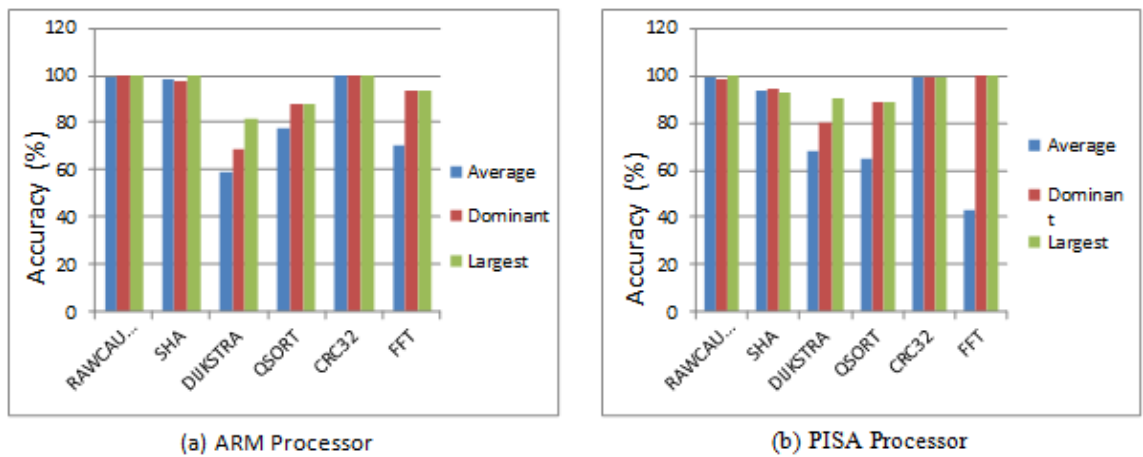

(a) ARM Processor  (b) PISA Processor

Figure 4 Accuracy in IPC Estimation for Various Block Length Criteria

Table 5 Cache Size as per Different Block Length Criteria

| Application | Cache Size Estimated (KB) | | | | | |
| --- | --- | --- | --- | --- | --- | --- |
| | ARM | | | PISA | | |
| | Avg. | Domi. | Largest | Avg. | Domi. | Largest |
| rawcaudio | 1 | 2 | 4 | 2 | 4 | 8 |
| sha | 2 | 4 | 8 | 4 | 8 | 16 |
| dijkstra | 1 | 2 | 8 | 2 | 4 | 16 |
| qsort | 4 | 8 | 8 | 8 | 16 | 16 |
| crc32 | 1 | 1 | 1 | 1 | 1 | 1 |
| fft | 8 | 64 | 64 | 16 | 128 | 128 |

## 5. DISCUSSION

The following observations have been noticed regarding the cache size estimation for optimal performance based on static analysis of the basic blocks proposed in this work.

(1) The error noticed in the estimation of the cache configuration is mainly because of the limitation of the LLVM tool. LLVM profiler cannot profile the library functions. Hence the basic blocks in the library functions are omitted from the basic block analysis done for cache configuration estimation. However, this is really not a limitation for embedded software in which most of the code is normally written by the developer rather than being called from library to optimize memory footprint and performance.

(2) In *qsort* application library function *qsort ()* is called which contributes to the major number of basic blocks. This leads to quite inaccurate result in estimation. The estimation accuracy is improved by including the library function into the application source code rather than calling from the library. This approach has increased the estimation accuracy by 11.80% for QSORT application.

(3) Cache line size and number of cache lines in SimpleScalar are always selected as an integer power of 2. Hence if the number of basic blocks in the original source code is near to boundary of power of 2 (i.e. 4, 8, 16, 32, 64,…). In such cases, more probability lies in exceeding the total blocks to the next higher boundary of power of 2 due to contributions from library functions. In such cases we can estimate the number of cache lines to next higher power of 2.

(4) The simulation time of SHA (small) application for different instruction cache size combinations varies from 7 to 23 seconds and for larger version it is average 200 seconds. In the proposed approach the cache size is estimated in time needed by LLVM profiler and the script to extract the total number of unique blocks and dominant block length. This time is less than 5 seconds in total for small as well as large version of the same application in MiBench. In this way the proposed approach offers fast exploration of the cache design space for given application and processor architecture.

(5) In certain applications, the cache line size estimated is significantly larger than the practically realizable. In such cases the line size can be reduced with proportional increment in number of cache lines without significant degradation in performance. For all applications with both processors the percentage change in IPC observed is within ±5% except DIJKSTRA (ARM).

## 6. CONCLUSION

In this paper, a technique for fast design space exploration of optimal instruction cache parameters (number of lines, line size) required for the given embedded software is presented. The proposed approach is validated using six benchmark applications from MiBench on two different processor architectures, ARM and PISA. The dominant block length based cache size estimation offers better cache-size versus performance trade-off compared to average and largest block lengths. The dominant block length based cache size offers average accuracy of 92.39% in IPC. In all cases, the IPC simulated for estimated cache size based on the dominant block is above 68% to that of the maximum IPC observed. This helps to avoid unnecessarily selecting a larger cache size. The methodology proposed is not tightly architecture dependent and can be easily adapted to other targets. The proposed technique estimates the cache parameters for optimal performance to achieve maximized IPC for source code with complete profiling possibility.

International Journal of Embedded Systems and Applications (IJESA) Vol.3, No.3, September 2013

## Authors

**Rajendra Patel** has received B.E. in Electronics and Communication Engineering in 2001 from the University of Bhavnagar, Gujarat, India. He has received M. Tech. (VLSI and Embedded System Design) in 2009 from MANIT, Bhopal, Madhya Pradesh, India. He is currently pursuing Ph.D. from the same institute. He had been accorded with distinction in both of his degrees. His research interest includes Embedded Software Performance, Hw/Sw Co-Design, Reconfigurable Systems and Digital System Design.

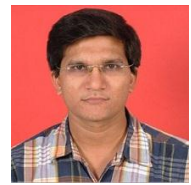

**Arvind Rajawat** has received his BE in Electronics & Communication Engineering from Government Engineering College, Ujjain, India and ME in Computer Engineering from SGSITS, Indore, India. He earned his PhD in the area of Communication Architecture Exploration in Codesign from MANIT, Bhopal, India. He has teaching experience of more than 20 years and he is currently working as Professor of Electronics and Communication Engineering at MANIT, Bhopal. His main areas of interest and research are Embedded System Design, Processor Architecture Exploration and Hardware Software Codesign.

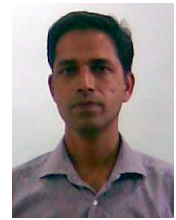